\documentclass[runningheads]{llncs}
\usepackage[T1]{fontenc}
\usepackage{svg}
\usepackage{graphicx}
\usepackage{multirow}
\usepackage{amssymb}
\usepackage{amsmath}
\usepackage{booktabs}
\usepackage{graphicx}
\usepackage{subcaption}
\usepackage{caption}
\usepackage{float}
\usepackage{multirow}
\begin{document}
\title{\underline{H}ierarchical \underline{S}patial-\underline{T}emporal \underline{G}raph-Enhanced Model for Map-\underline{Match}ing}
%
%
\author{Anjun Gao\thanks{Equal contribution.}\inst{1} \and
Zhenglin Wan$^*$\inst{2} \and
Pingfu Chao\thanks{Corresponding author.}\inst{1} \and 
Shunyu Yao \inst{3}}

\authorrunning{Anjun et al.}

\institute{School of Computer Science and Technology, Soochow University, Suzhou, China \\
\email{kalis1e7@gmail.com, pfchao@suda.edu.cn} \and
Metasequoia Intelligence, Shenzhen, China \\
\email{carlos@metaseq.ai} \and
Department of Astronomy, The Ohio State University, Columbus, OH 43202 USA \\
\email{yao.869@buckeyemail.osu.edu}}
\maketitle              
\begin{abstract}
The integration of GNSS data into portable devices has led to the generation of vast amounts of trajectory data, which is crucial for applications such as map-matching. To tackle the limitations of rule-based methods, recent works in deep learning for trajectory-related tasks occur. However, existing models remain challenging due to issues such as the difficulty of large-scale data labeling, ineffective modeling of spatial-temporal relationships, and discrepancies between training and test data distributions. To tackle these challenges, we propose \textbf{HSTGMatch}, a novel model designed to enhance map-matching performance. Our approach involves a two-stage process: hierarchical self-supervised learning and spatial-temporal supervised learning. We introduce a hierarchical trajectory representation, leveraging both grid cells and geographic tuples to capture moving patterns effectively. The model constructs an \textit{Adaptive Trajectory Adjacency Graph} to dynamically capture spatial relationships, optimizing GATs for improved efficiency. Furthermore, we incorporate a \textbf{Spatial-Temporal Factor} to extract relevant features and employ a decay coefficient to address variations in trajectory length. Our extensive experiments demonstrate the model's superior performance, module effectiveness, and robustness, providing a promising solution for overcoming the existing limitations in map-matching applications. The source code of HSTGMatch is publicly available on GitHub at \url{https://github.com/Nerooo-g/HSTGMatch}.

\keywords{Map-matching \and Spatial-temporal trajectories \and Machine learning.}
\end{abstract}

\section{Introduction}
In today's world, the Global Navigation Satellite System (GNSS) is deeply integrated into all types of portable devices, generating a continuous stream of trajectory data that fuels a variety of applications. Traditional rule-based methods \cite{Goh2012onlineHMM,Huang2018frequent,Newson2009HMM,wei2013Frechet} have limitations that fail to adapt to complex road networks and poor robustness to the noisy faces challenges in some cases. Recently, there has been significant interest in leveraging deep learning to enhance trajectory-related tasks, such as map-matching \cite{feng2020deepmm,jin2022transformer,liu2023graphmm}. However, the practical application of deep learning techniques in map-matching is hindered by the need for large, high-quality labeled datasets. This challenge is compounded by three main issues:

\begin{itemize}
\item \textbf{Difficulty in large-scale data labeling}: Labeling extensive trajectory datasets is particularly challenging due to inherent inaccuracies. These inaccuracies arise from positioning errors and uncertainties within the road network, making trajectory data labeling more complex than in other fields.
\item \textbf{Ineffective modeling of spatial-temporal relationships}: Many models simply adapt architectures from other fields, such as natural language processing, without tailoring them to the specific requirements of the map-matching task.
\item \textbf{Discrepancies between training and test data distributions}: Existing methods struggle to create data structures that accommodate variations between training and test data distributions, resulting in inconsistent performance when applied to real-world scenarios.
\end{itemize}

To address these challenges, existing methods focus on optimizing trajectory representation from spatial-temporal perspectives. One common approach involves rasterizing a map into grids and mapping GPS points to corresponding grid cells \cite{feng2020deepmm}. However, using one-hot encoding for grid cells fails to capture spatial relationships between adjacent grids. To overcome this, GraphMM \cite{liu2023graphmm} introduces a trajectory graph to model these spatial relationships. Additionally, to ensure model convergence and work with limited labeled data, the grid granularity should not be too fine. However, a coarse grid size may overlook variations between different locations, significantly impairing the performance of high-precision tasks like map-matching.

Despite these advancements, several issues remain. The lack of high-quality trajectory data often results in poor performance during testing. Although previous work \cite{feng2020deepmm} suggests data augmentation, the generated data can still differ significantly from ground-truth data due to various movement patterns. Minimizing performance loss caused by distribution differences between training and test sets remains a priority. Furthermore, explicitly incorporating spatial-temporal factors into models is crucial for improving performance.

We propose a model named \textbf{HSTGMatch} to address the challenges in map-matching and enhance existing solutions. Inspired by many self-supervised tasks \cite{DBLP:conf/nips/TungTYF17,DBLP:conf/cvpr/SermanetLHL17}, this model operates in two stages: hierarchical self-supervised learning and spatial-temporal supervised learning. In the self-supervised learning phase, it aims to capture spatial features within trajectories. To accurately capture moving patterns while balancing convergence, we employ a hierarchical representation that utilizes both grids (cells) and scaled tuples (\textit{longitude}, \textit{latitude}). We then construct an \textit{Adaptive Trajectory Adjacency Graph} to model spatial relationships among grids, assigning weights to edges to adapt across different datasets. This approach maintains stability, as constructed edges are unaffected by dataset distribution, while dynamically capturing variations among datasets. Additionally, we optimize Graph Attention Networks (GATs) \cite{veličković2018graph} to simplify operations, enhancing efficiency and training speed.

During downstream task training, we incorporate the \textbf{Spatial-Temporal Factor} into embeddings to explicitly extract features. We also address the relevance variation with trajectory length by introducing a decay coefficient.

In summary, our main contributions are:

\begin{itemize}
\item We introduce \textbf{HSTGMatch}, a novel model for the map-matching task. Our model leverages a self-supervised architecture and an innovative trajectory representation, incorporating spatial-temporal elements to achieve exceptional performance.
\item We propose a \textbf{Hierarchical Self-supervised Learning} for trajectory representation. Our model input includes both fine-grained and coarse-grained trajectory features, using optimized GATs to model spatial relationships, balancing stability and adaptability.
\item We introduce the \textbf{Spatial-Temporal Factor} to explicitly extract spatial-temporal features from trajectories and employ a decay coefficient to manage long-range dependency issues.
\item We conduct extensive experiments to validate the model’s overall performance, module effectiveness, and robustness.
\end{itemize}


\section{Related Works}
\subsection{Traditional Methodologies}

Traditional methodologies are predominantly rule-based. For instance, the Fréchet method \cite{wei2013Frechet} employs the Fréchet distance between trajectories and road maps to determine the optimal route. The FPM approach \cite{Huang2018frequent} relies on historical similarity within the similarity model, while HMM \cite{Newson2009HMM} is a widely used and effective candidate-graph-like model. However, these traditional methods have two main shortcomings: they often fail to capture the potential movement patterns in complex road networks, leading to performance limitations under certain conditions, and they exhibit poor robustness to noise, making them heavily reliant on high-quality data.

\subsection{Machine Learning Models}
Recent efforts, such as DeepMM \cite{feng2020deepmm}, have introduced stacked bidirectional LSTM layers and attention mechanisms \cite{vaswani2017attention} to better analyze spatial patterns in trajectories. However, these models struggle to extract temporal features, particularly for longer sequences, due to the inherent limitations of RNN architectures. To overcome these challenges, the Transformer-based Map-Matching Model \cite{jin2022transformer} incorporates self-attention mechanisms to enhance performance. Additionally, GraphMM employs a graph-based approach that explicitly leverages spatial correlations, effectively utilizing both road and trajectory graph topologies to align road segments and trajectories in latent space.

\section{Preliminaries}

\subsection{Problem Definitions}
\begin{definition}[\textbf{Trajectory}]\label{trajectory}
    A trajectory $T$ is a series of chronologically ordered points $T=\left<p_{1} \rightarrow p_{2} \rightarrow \dots \rightarrow p_{n}\right>$ representing the historical trace of an object. Each point $p_{i}=\left<x,y,t\right>$ indicates the location \textit{(longitude, latitude)} of the object at time $p_{i}.t$.
\end{definition}

\begin{definition}[\textbf{Grid}]\label{grid}
    A map region can be partitioned into $m \times n$ grids of equal size through horizontal and vertical lines. Each grid cell is assigned a unique ID, i.e. a grid so that any GPS point $p_j$ maps to one and only one grid area with a certain ID.
\end{definition}

\begin{definition}[\textbf{Distance Interval}]\label{distance_inerval}
    Given a trajectory $T$, a distance interval is represented as the absolute difference in geographical values between a point $p_{i}$ from the first one $p_{0}$. The initial distance interval is zero.
\end{definition}

\begin{definition}[\textbf{Time Interval}]\label{time_interval}
    Given a trajectory $T$, a time interval is represented as the absolute difference in timestamps between a point $p_{i}$ from the first one $p_{0}$. The initial time interval is zero.
\end{definition}

\begin{definition}[\textbf{Segment-based Route}]\label{route}
    A segment-based route $R$ can be expressed as $R=<c_1,c_2,...,c_i>$, where $c_m$ represents a road identified by a unique ID on the map.
\end{definition}

\begin{definition}[\textbf{Map-Matching}]\label{map_matching}
    Map-matching is the process by which a given trajectory $T$, composed of a sequence of GPS points, is aligned with the most accurate corresponding sequence of roads to form a segment-based route $R$.
\end{definition}

\begin{figure*}[htbp]
  \centering
  \includegraphics[width=1\textwidth]{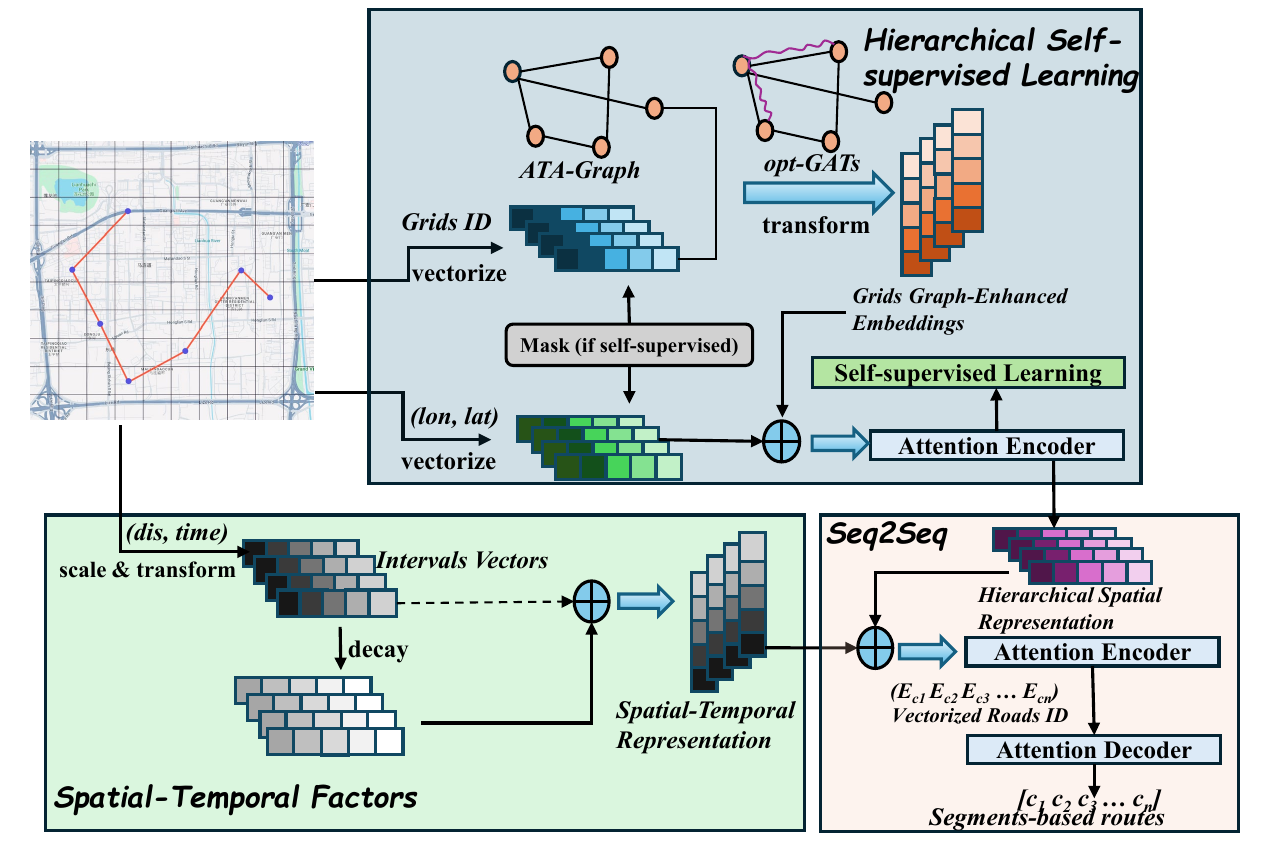}
  \caption{HSTGMatch architecture overview, $\oplus$: concatenate}
\label{fig1}
\end{figure*}

\subsection{Model Overview}
As illustrated in Fig \ref{fig1}, the workflow of \textbf{HSTGMatch} is presented, comprising three main modules: \textbf{Hierarchical Self-supervised Learning}, \textbf{Spatial-Temporal Factors}, and the \textbf{Seq2Seq Model}. Initially, we divide the maps into several grids and assign a unique number to each grid. Each trajectory point is then mapped to its corresponding grid, and coordinates are scaled accordingly. During the self-supervised learning phase, the model focuses on decoding masked points. After several epochs, once the loss stabilizes, we transfer the architecture and integrate the Spatial-Temporal Factors module for end-to-end supervised training. During training, vectorized Road IDs are incorporated into the decoding process to facilitate teacher-forcing training. In the inference stage, the Decoder generates the corresponding routes based on the probability distribution of the final vectors.

\section{Methodology}


\subsection{Graph Embedding}
\subsubsection{Constructing Adaptive Trajectory Adjacent Graph}
First, we map each trajectory point onto a grid based on the spatial range. A previous work \cite{liu2023graphmm} constructs a trajectory graph by considering the distribution across the entire map, assigning higher weights to edges connecting grids that frequently contain successive trajectory points. However, this approach requires the training and test dataset's distribution to be consistent.

Given that the map-matching process is highly dependent on spatial features, we introduce a distance threshold for the grids. Two grids can form an edge if and only if their distance is less than this threshold. To dynamically adapt to different datasets, we assign weights to the graph based on the total number of trajectory points within each grid. This results in the definition of the Adaptive Trajectory Adjacency Graph $\mathcal{G}$.
\begin{definition}[\textbf{Adaptive Trajectory Adjacent Graph}] \label{ATAG}
    An Adaptive Trajectory Adjacency Graph (ATA-Graph) is represented as $\mathcal{G_T}=(V_T,E_T,W_T)$, where $V_T$ denotes the nodes (grids) in the graph, $E_T$ denotes the edges formed based on geographical proximity, and $W_T$ denotes the weights calculated from the distribution of trajectory points.
\end{definition}

\subsubsection{Optimized Graph Attention Networks} GATs \cite{veličković2018graph} offer a flexible and adaptive approach to processing graph-structured data by leveraging attention mechanisms. This adaptability is particularly advantageous for map-matching, where handling complex spatial relationships and noisy data is crucial. Inspired by the works \cite{he2020lightgcn,rao2022graph} that simplify and remove non-linear activation for GCNs \cite{kipf2017semisupervised}, we decide to employ optimized GATs (opt-GATs) to aggregate features. To improve efficiency and training speed, we simplified the calculation of attention scores by discarding the concatenation of features of nodes and their neighbors. Instead, we directly apply the dot product between the features of two nodes and utilize a factor $\frac{1}{\sqrt{d}}$ and Softmax function to scale the attention scores. This can be represented as:
\begin{equation}
    \alpha_{ij} = \frac{\exp(\frac{1}{\sqrt{d}}\mathbf{W}h_i\cdot \mathbf{W}h_j)}{\sum_{m \in \mathcal{N}_i}\exp(\frac{1}{\sqrt{d}}\mathbf{W}h_i\cdot \mathbf{W}h_m)},
\end{equation}
where $\alpha_{ij}$ represents the attention coefficient between node $i$ and node $j$. The dimensionality of the node features is denoted by $d$. The weight matrix applied to the node features is represented by $\mathbf{W}$. The feature vectors of node $i$ and node $j$ are denoted by $h_i$ and $h_j$, respectively. The set of neighbors of node $i$ is indicated as $\mathcal{N}_i$.

To better adapt to variations across different datasets, it is significant to aggregate information about the data distribution. Upon applying the multi-head attention, the updated representation for grid $i$ is obtained by first aggregating the attention-weighted features from its neighborhood  $\mathcal{N}_i$  across all $K$ heads. Each head k computes an attention score $ \alpha_{ij}^k$ for the grid $i$ and its neighbor grid $j$, which is then used to scale the transformed features $\mathbf{W}^kh_j$. The transformed and attention-scaled feature vectors from all heads are then concatenated and averaged. Finally, the result incorporates the transformed weighted vectors to yield the final updated feature $h_i'$ for the grid $i$, where $W_l$ denotes the matrix that transforms the $h_j$, and $\gamma_j$ represents the weights of neighbor $j$ when constructing the \textit{ATA-Graph} in Definition \ref{ATAG}. The equation summarizing this process is given as:
\begin{equation}
    h_{i}^{\prime}=\frac{1}{K}\sum_{k=1}^{K}\sum_{j\in \mathcal{N}_{i}}\alpha_{ij}^{k}\mathbf{W}^{k}h_{j}\oplus\sum_{j\in \mathcal{N}_{i}} W_l h_{j} \gamma_j
\end{equation}

\subsection{Hierarchical Self-supervised Learning}
The self-supervised learning has proved to be powerful and effective in many downstream tasks in different fields. In the scenario of trajectories, capturing spatial and temporal relationships within each trajectory is significant. Besides, the attention mechanism \cite{vaswani2017attention} is expert at dealing with variable length sequences compared to RNN models, thus, we utilize the attention-based encoder to pre-train the dataset.

\subsubsection{Unsupervised Objectives}
Unlike in natural language where semantic meanings can drastically change between two tokens, spatial-temporal tokens within a trajectory exhibit more consistent properties. Therefore, our approach involves masking spans of tokens to better enable the model to grasp these long-range dependencies. The T5 \cite{raffel2020exploring} approach uses a replace spans strategy, where masked spans are substituted with unique sentinel tokens. We, however, introduce a modification for trajectory data: employing a shared sentinel token for all masked spans and only keeping the encoder to improve efficiency instead of seq2seq models. This promotes the learning of more generalized representations, as opposed to position-specific ones. By utilizing a shared sentinel token, the model is encouraged to generalize better, resulting in faster convergence and enhanced stability during training, without the complexity of managing numerous sentinel tokens.

\subsubsection{Optimization}
In the self-supervised learning phase, models should aim to capture both the broad and nuanced variations among trajectories that represent hierarchical attributes with particular attention to spatial features. Consequently, our approach involves masking grids and $(longitude, latitude)$ tuples according to the rules defined by our unsupervised objectives in the encoder. After encoding, the resultant hidden vectors $h_s \in \mathbb{R}^{L\times D}$ are fed through two distinct fully connected layers to reconstruct the masked values. The optimization function focuses primarily on the accuracy of the predicted unmasked values. Formally, we can give the formula as,
\begin{equation}
    L(y^{u}_{g_i},\hat{y}^{u}_{g_i},y^{u}_{p_i},\hat{y}^{u}_{p_i}) = k\cdot[ - \sum_{u \in \mathcal{U}} \sum_{i=1}^{n} y^{u}_{g_i} \log(\hat{y}^{u}_{g_i})]+(1-k)\cdot \sum_{u \in \mathcal{U}}\sqrt{\frac{1}{n} \sum_{i=1}^n (y^{u}_{p_i} - \hat{y}^{u}_{p_i})^2},
\end{equation}
where $k$ is a hyper-parameter that balances the contribution of two separate tasks: grid prediction and tuple prediction. The terms $y^{u}_{g_i}$ and $y^{u}_{p_i}$ represent the true labels of the i-th grid and the tuple of $(longitude, latitude)$ for trajectory $u$ respectively. This balanced approach ensures that the model effectively learns to predict spatial features while being robust to variations in the input data.

\subsection{Supervised Learning}

\subsubsection{Spatial-Temporal Factor}
In map-matching tasks, effectively capturing spatial-temporal features among trajectories is crucial for accurately generating road paths. To address this, we explicitly consider both geographical distance and time intervals. Inspired by the HST-LSTM \cite{kong2018hst}, we transform and scale these distance and time intervals using the following equation,
\begin{equation}
r = \frac{W_{u(v_r)}[u(v_r) - v_r] + W_{l(v_r)}[v_r - l(v_r)]}{u(v_r) - l(v_r)},
\end{equation}
where $r \in \mathbb{R}^D$ stands for the transformed vector of an interval and $v_r$ denotes the absolute value of intervals for sequences. The $u(v_r)$ and $l(v_r)$ represent the upper and lower bound values of a slot concerning time or distance interval sequences. $W \in \mathbb{R}^{N_r \times D}$ denotes the spatial or temporal factor matrix, where $N_r \in R$ is the total number of time or distance slots, respectively. Since both time and distance intervals represent absolute value differences from the first point, their importance diminishes as the sequence length increases. For instance, the last point in a trajectory is nearly irrelevant to the first point because the corresponding road of a point depends on only a few nearby context points. We incorporate this decay in importance using the following operations,
\begin{equation}
    r'= \frac{\log(v_r)}{s}W_ir,
\end{equation}
\begin{equation} \label{eq6}
    r''=r \oplus r',
\end{equation}
where $r'$ denotes the coefficient that determines the relevance of the s-th position token to the first token, $W_i$ denotes the matrix transforming $r$, and $r''$ denotes the final representation of distance or time interval for $v_r$.

\subsubsection{Training}
Self-supervised learning makes the model understand the hierarchical spatial correlation, which is also significant for the map-matching task in supervised learning. Therefore, the final prediction layer in self-supervised learning is discarded and multiply a matrix $W_s \in \mathbb{R}^{D\times D'}$ to transform the hidden vector $h_s$ into $h_s' \in \mathbb{R}^{L\times D'}$. The training process can be formulated as,
\begin{equation}
    E=\text{Encoder}(W'[W_sh_s\oplus s \oplus t]),
    \end{equation}
\begin{equation}
    E'=\text{Decoder}(E,R),
\end{equation}
\begin{equation}
    Z=E'W_q^{\top}.
\end{equation}
To obtain the probability distribution over all possible roads, we apply the softmax function:
\begin{equation}
    P(y = c \mid h_s, s, t, R) = \frac{\exp(Z_{l}^c)}{\sum_{c'=1}^C \exp(Z_{l}^{c'})} \quad \text{for } l \in \{1, \ldots, L\}, \; c \in \{1, \ldots, C\},
\end{equation}
where $P$ denotes the probability of selecting road $c$ given the hidden vector $h_s$, $s$ and $t$ denote derived spatial and temporal vectors by Equation \ref{eq6} respectively, $W' \in \mathbb{R}^{D_f \times D_k}$ denotes the matrix projects concatenated vectors. $L$ denotes the length of a trajectory, $R$ denotes the route defined in Definition \ref{route}, $W_q \in \mathbb{R}^{C \times D_f}$ denotes the linear matrix, and $C$ denotes the total number of roads.

\subsubsection{Learning}
In the optimization phase, we aim to minimize the loss function, which quantifies the discrepancy between the predicted probability distributions and the actual road labels. For the map-matching task, we use the multi-class cross-entropy loss function, which is well-suited for classification problems involving probabilistic outputs. Finally, the loss function can be formulated as,
\begin{equation}
\mathcal{L} = - \sum_{l=1}^{L} \sum_{c \in C} y_{l}^c \log(\hat{y}_{l}^c),
\end{equation}
where $l$ and $c$ are defined above.

\section{Experiments}

\subsection{Datasets}
To validate the effectiveness of our model across datasets of varying scales, we constructed datasets from three areas in Beijing, each with different dimensions: approximately $5km \times 5km$, $8km \times 8km$, and $12km \times 12km$. These regions are situated near the city center and feature complex road networks. The datasets comprise a total of $33,690$, $79,543$, and $193,272$ vehicle trajectories, respectively, collected at random sampling intervals. For our experiments, we divided each dataset into training and testing subsets, using the initial $80\%$ of the trajectories for training and the remaining $20\%$ for testing.

\subsection{Baselines}

\begin{itemize}
    \item LSTM \cite{hochreiter1997long}: This traditional RNN architecture incorporates forget gates and memory cells to capture vital information in sequential data efficiently.
    \item ST-RNN \cite{liu2016predicting}: The ST-RNN model employs time-specific and distance-specific transition matrices to capture the temporal cyclic effects and geographical influences respectively.
    \item HST-LSTM \cite{kong2018hst}: HST-LSTM integrates spatial and temporal influences within an LSTM framework to address data sparsity, enhanced by a hierarchical extension.
    \item DeepMM \cite{feng2020deepmm}: DeepMM employs a stacked and bidirectional LSTM \cite{hochreiter1997long} with an attention mechanism \cite{luong2015effective} to capture the spatial information of trajectories, which are represented as grids.
    \item Transformer-based Map-matching Model \cite{jin2022transformer}: it adopts the standard Transformer architecture. It utilizes normalized longitudes and latitudes to represent trajectories, and also incorporates temporal representation.
    \item GraphMM \cite{liu2023graphmm}: A graph-based approach that explicitly leverages all the aforementioned correlations. This method exploits the inherent graph structure of map matching by incorporating graph neural networks and conditional models.
\end{itemize}

\subsection{Evaluation Criteria}
To evaluate the effectiveness of our map-matching algorithm, we employ precision, recall, and F1-score as key performance metrics. These are defined as follows:
\begin{equation}
\text{Precision} = \frac{\text{len}(P^{m} \cap P^{g})}{\text{len}(P^{m})},
\end{equation}

\begin{equation}
\text{Recall} = \frac{\text{len}(P^{m} \cap P^{g})}{\text{len}(P^{g})},
\end{equation}

\begin{equation}
\text{F1-Score} = 2 \times \frac{\text{Precision} \times \text{Recall}}{\text{Precision} + \text{Recall}},
\end{equation}
where $P^{m}$ represents the matched routes while $P^{g}$ denotes the ground truth routes. Both are segment-based routes which are defined in Definition \ref{route}. The function $\text{len}()$ calculates the length of a route, and $P^{m} \cap P^{g}$ refers to the set of correctly matched road segments.

\subsection{Settings}
The map is divided into $100\times100$ meter grids, and the coordinates and time intervals are normalized using Z-score. The model has a d\_model of 128, and both the encoder and decoder consist of 4 layers. The attention mechanism employs 8 attention heads, each with 128 dimensions. For training and evaluating these tasks, we utilize 2 NVIDIA Tesla A100 GPUs and an Intel Xeon Gold 5220R CPU.

\begin{table*}[htbp]

\caption{Overall Performances}
\label{table1}
\centering
\setlength{\tabcolsep}{4pt} 
\begin{tabular}{|c|c|c|c|c|c|c|c|c|c|}
\hline
\multirow{2}{*}{Models} & \multicolumn{3}{c|}{Beijing-S} & \multicolumn{3}{c|}{Beijing-M} & \multicolumn{3}{c|}{Beijing-L} \\
\cline{2-10}
& \it P(\%) & \it R(\%) & \it F1(\%) & \it P(\%) & \it R(\%) & \it F1(\%)& \it P(\%) & \it R(\%) & \it F1(\%) \\
\hline
LSTM & 63.21 & 60.19 & 61.66 & 60.75 & 59.34 & 60.04 & 50.15 & 49.28 & 49.71 \\
ST-RNN & 65.14 & 61.08 & 63.04 & 62.53 & 60.94 & 61.72 & 52.13 & 53.87 & 52.99 \\
HST-LSTM & 70.47 & 71.22 & 70.84 & 73.27 & 75.59 & 74.41 & 56.36 & 57.12 & 56.74 \\
DeepMM & 74.24 & 73.35 & 73.79 & 78.69 & 81.47 & 80.06 & 59.03 & 60.42 & 59.72 \\
TransMM$^{\mathrm{*}}$ & 80.18 & 79.47 & 79.82 & 85.19 & 86.16 & 85.67 & 77.26 & 79.10 & 78.17 \\
GraphMM & 84.95 & 84.70 & 84.82 & 87.78 & 88.83 & 88.30 & 84.52 & 85.17 & 84.84 \\
\hline
\textbf{HSTGMatch} & \textbf{88.16} & \textbf{89.03} & \textbf{88.59} & \textbf{90.58} & \textbf{90.27} & \textbf{90.42} & \textbf{88.94} & \textbf{89.50} & \textbf{89.22} \\
\hline
\multicolumn{10}{l}{$^{\mathrm{*}}$Denotes Transformer-based Map-matching Model.}

\end{tabular}
\end{table*}

\subsection{Overall Performances}
As Table \ref{table1} illustrates, we can see the differences in performances on different datasets. For this task, the progression from LSTM-based models to Transformer and graph-based architectures highlights significant advancements in map-matching capabilities. The LSTM model serves as a foundational baseline, yet its limited ability to capture long-range dependencies results in moderate performance across all datasets. The introduction of hierarchical and spatial-temporal architectures, such as HST-LSTM, demonstrates a marked improvement, particularly on medium-sized datasets, indicating the value of capturing complex spatial-temporal patterns. DeepMM, building on LSTM's strengths and introducing the attention mechanism, shows enhanced performance but still falls short compared to Transformer-based approaches. TransMM, leveraging the self-attention mechanism, excels in capturing spatial dependencies, achieving superior performances compared to its predecessors. GraphMM further advances performance by integrating structural information, underscoring the benefits of a graph-based approach. HSTGMatch, our proposed model, surpasses all existing models, achieving exceptional performances across all datasets. This success can be attributed to the model's ability to effectively integrate hierarchical self-supervised learning and the spatial-temporal factor along with enhanced graph representations, providing a comprehensive understanding of trajectory data. The overall modest improvement across the various metrics suggests that the accuracy of the matching process may have reached a high plateau, and disparities in trajectory coverage due to the reliance on road networks remain a challenge in certain areas.

\begin{figure}[H]
    \centering
    
    \begin{subfigure}[b]{0.49\textwidth}
        \centering
        \includegraphics[width=\textwidth]{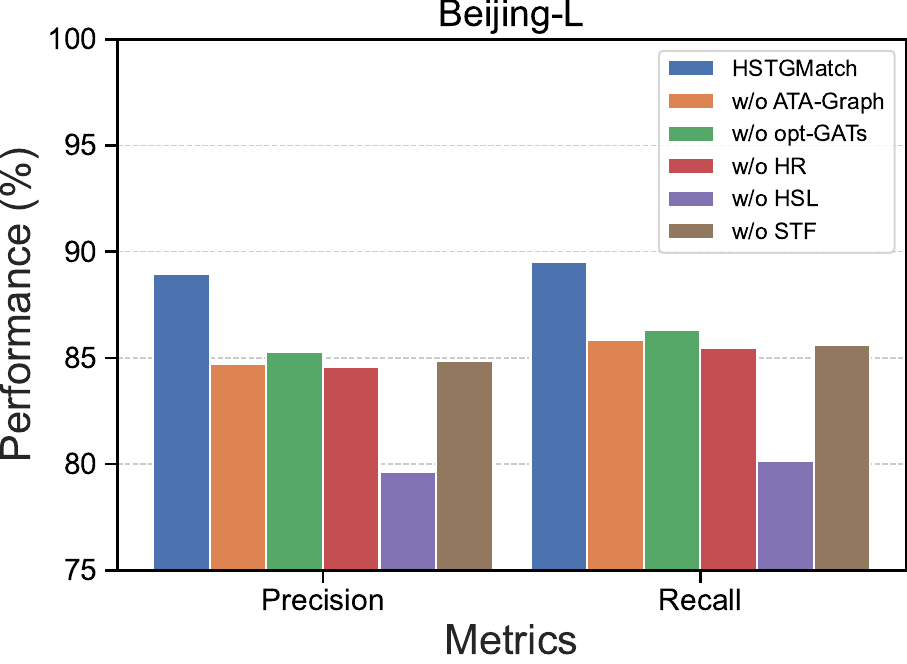}
        \caption{Beijing-L}
        \label{fig:beijing_l}
    \end{subfigure}
    \hfill
    \begin{subfigure}[b]{0.49\textwidth}
        \centering
        \includegraphics[width=\textwidth]{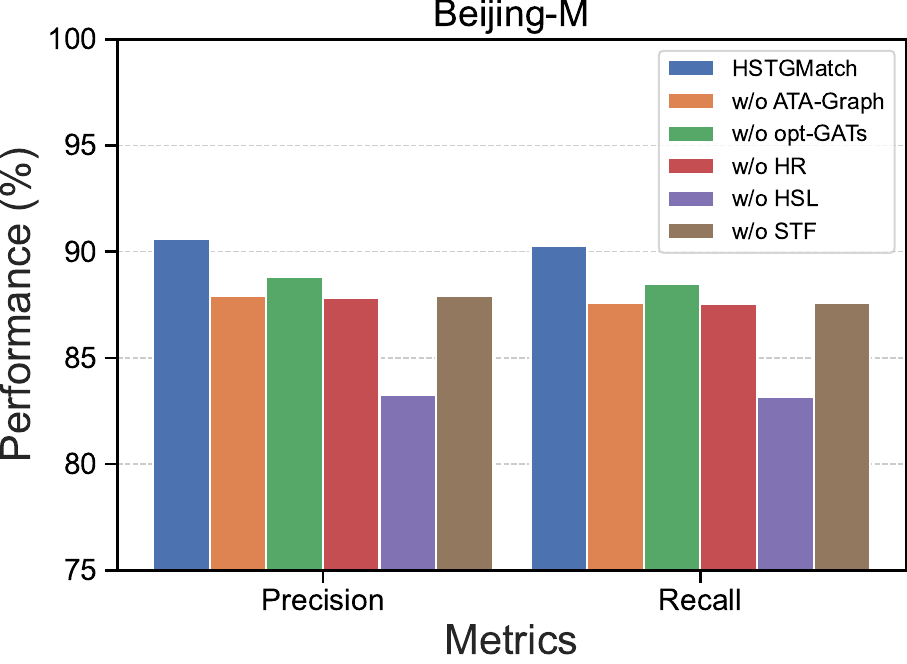}
        \caption{Beijing-M}
        \label{fig:beijing_m}
    \end{subfigure}
    \hfill

    \caption{Performance on Beijing datasets (Precision and Recall). Each subfigure represents a different dataset; w/o opt-GATs denotes replacing GCNs instead; HR denotes Hierarchical Representation; HSL denotes Hierarchical Self-supervised Learning; STF denotes Spatial-Temporal Factors.}
    \label{fig: ablation}
\end{figure}

\subsection{Ablation Study}
As illustrated in Fig \ref{fig: ablation}, the effectiveness of our modules is clearly demonstrated. The complete \textbf{HSTGMatch} model consistently achieves the highest precision and recall across both datasets. Replacing opt-GATs with GCNs also results in decreased performance, underscoring the importance of attention mechanisms for capturing complex relationships. Notably, handling larger maps presents a challenge due to the increased number of grids that need to be managed and processed. Our graph-based designs are particularly effective in this context, as they enable the efficient handling of complex spatial information. Without hierarchical representation, precision and recall decrease, suggesting that capturing multi-level features is crucial for performance. We also observe that \textbf{Spatial-Temporal Factors} become increasingly important with larger datasets. Moreover, the importance of self-supervised learning is highlighted, as it empowers the model to more explicitly grasp complex features within trajectories, enhancing its ability to make accurate predictions without the need for extensive labeled data.

\begin{figure}[H]
    \centering
    
    \begin{subfigure}[b]{0.49\textwidth}
        \centering
        \includegraphics[width=\textwidth]{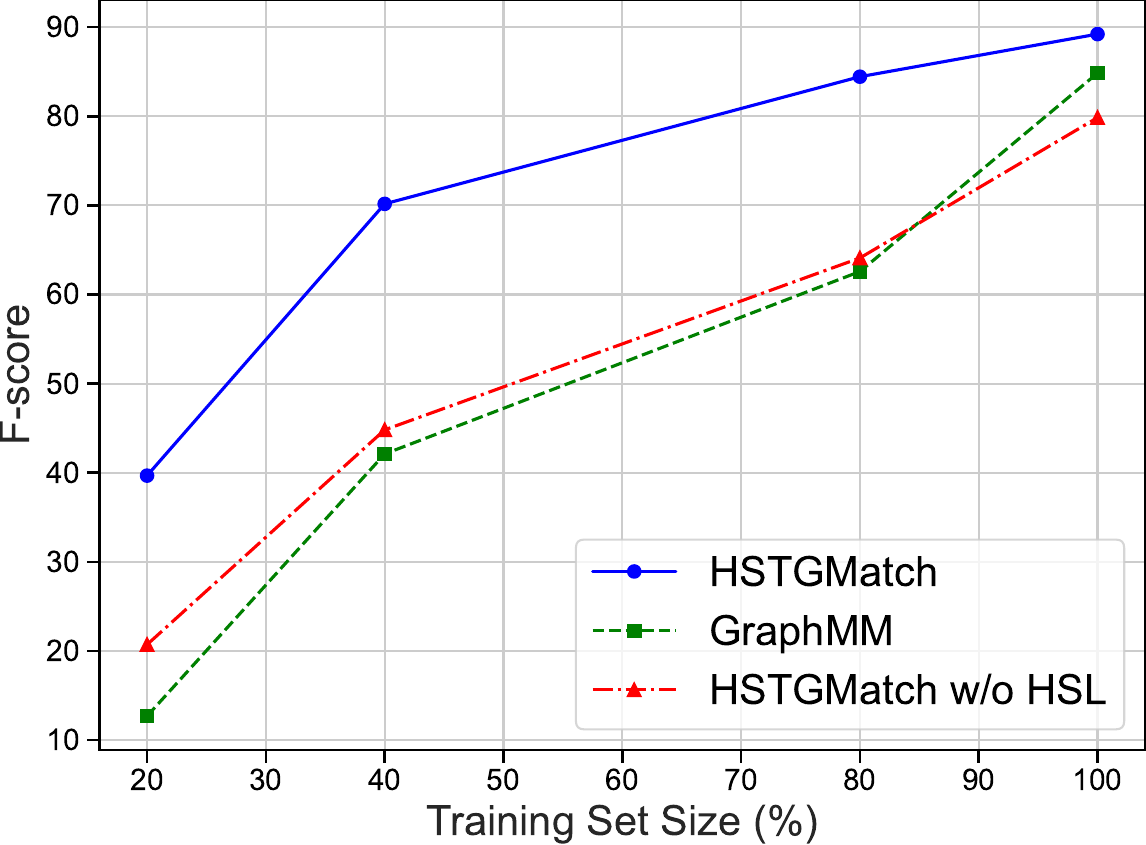}
        \caption{Beijing-L}
        \label{fig:beijing_l2}
    \end{subfigure}
    \hfill
    \begin{subfigure}[b]{0.49\textwidth}
        \centering
        \includegraphics[width=\textwidth]{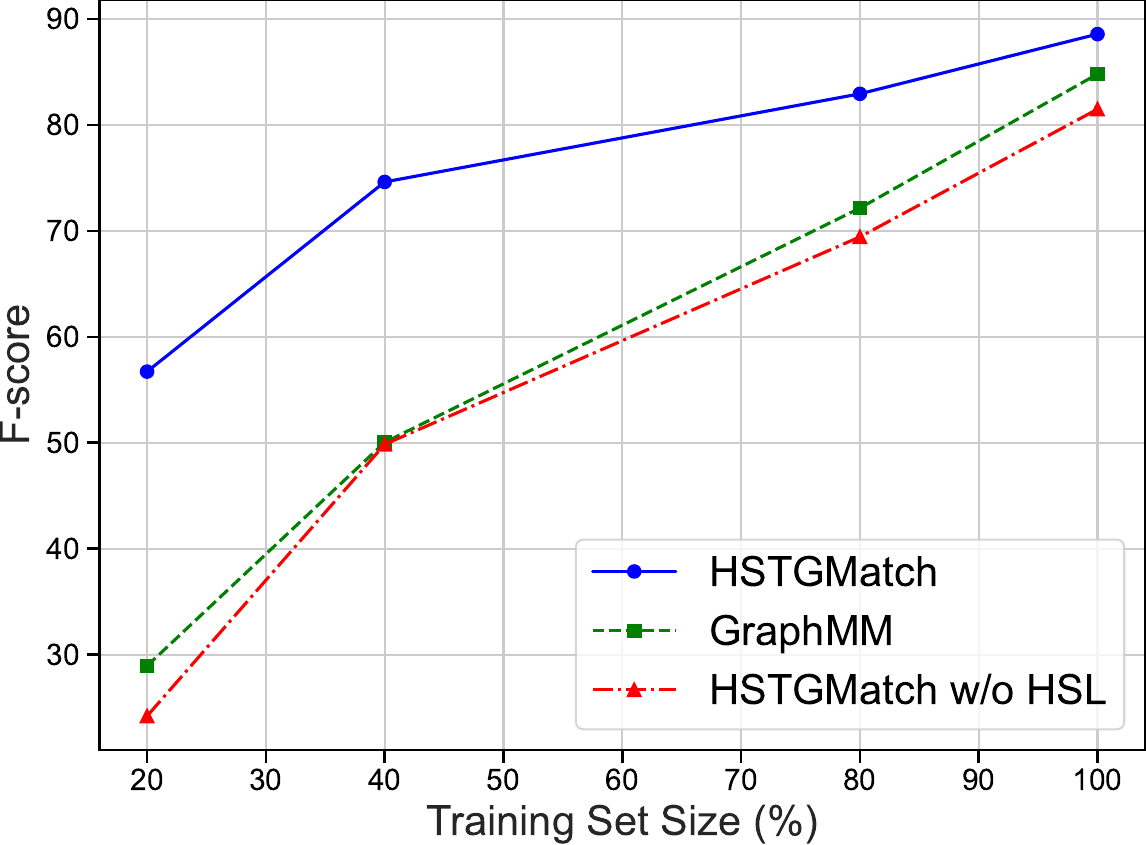}
        \caption{Beijing-S}
        \label{fig:beijing_s}
    \end{subfigure}
    \hfill
    
    \caption{Robustness Study. Each subfigure represents a different dataset.}
    \label{fig: robustness}
\end{figure}

\subsection{Robustness Study}
As illustrated in Fig \ref{fig: robustness}, we validate the robustness of our model, demonstrating how \textbf{Hierarchical Self-supervised Learning} significantly enhances the representation of trajectories. Notably, even when the training dataset is reduced to 40\% of its original size, the performance reduction remains substantial. The self-supervised learning approach allows the model to effectively leverage unlabeled data, thereby improving its ability to model spatial relationships. This capability is particularly advantageous in scenarios where labeled data is scarce or costly to obtain. Moreover, in large-scale maps, we observe that models lacking self-supervised learning are highly dependent on having sufficient data. This further supports the conclusion that our model improves the modeling of spatial representation through self-supervised learning.

\section{Conclusion and Discussion}
In this paper, we presented \textbf{HSTGMatch}, an innovative model for enhancing map-matching performance using trajectory data. By integrating self-supervised and spatial-temporal supervised learning, our approach effectively captures moving patterns through hierarchical trajectory representation. The model's \textbf{Hierarchical Self-supervised Learning} and \textbf{Spatial-Temporal Factor} address key challenges such as spatial relationships and trajectory length variations. Our experiments confirm the model's superior performance and robustness over existing methods, providing a promising solution to overcome limitations in map-matching applications.

Future work could explore the use of very large-scale unlabeled data to develop a pre-training model adaptable to various downstream tasks. Additionally, further optimization of trajectory representation might involve enhancing the model's ability to transfer across different regions.

%
%
%
\bibliographystyle{splncs04}
\bibliography{mybibliography}
%




\end{document}